\pdfoutput=1

\documentclass{aa}

\usepackage{graphicx}
\usepackage[varg]{txfonts}
\usepackage{amsmath}
\usepackage{verbatim}
\usepackage{latexsym}
\usepackage{multirow}
\usepackage{fixltx2e}
\usepackage{color}
\usepackage{rotating}

\begin{document}

\title{Similarities of magnetoconvection in the umbra and in the penumbra of sunspots}

\author{B. L\"optien\inst{1}
\and A. Lagg\inst{1,3}
\and M. van Noort\inst{1}
\and S.~K. Solanki\inst{1,2}}

\institute{Max-Planck-Institut f\"ur Sonnensystemforschung, Justus-von-Liebig-Weg 3, 37077 G\"ottingen, Germany
\and School of Space Research, Kyung Hee University, Yongin, Gyeonggi, 446-701, Republic of Korea
\and Department of Computer Science, Aalto University, PO Box 15400, FI-00076 Aalto, Finland}

\date{Received <date> /
Accepted <date>}

\abstract {It is unclear why there is a rather sharp boundary in sunspots between the umbra and the penumbra. Both regions exhibit magnetoconvection, manifesting in penumbral filaments in the penumbra and in umbral dots in the umbra.}
{Here we compare the physical properties of umbral dots and penumbral filaments. Our goal is to understand how the properties of these convective features change across the boundary between the umbra and the penumbra and how this is related to the rapid increase in brightness at the umbra-penumbra boundary.}
{We derived ensemble averages of the physical properties of different types of convective features based on observations of two sunspots with Hinode.}
{There are strong similarities between the convective features in the outer parts of the umbra and the ones in the penumbra, with most physical parameters being smooth and continuous functions of the length of the features.}
{Our results indicate that the transition in brightness from the umbra to the penumbra is solely caused by an increased effectiveness of magnetoconvection within individual convective cells. There is no significant difference in the number density of convective elements between the outer umbra and the inner penumbra. Penumbral filaments exhibit a larger area and a higher brightness compared to umbral dots. It is still unclear, how exactly the underlying magnetic field causes the increase in size and brightness of convective features in the penumbra.}


\maketitle

\section{Introduction}
Sunspots consist of a dark umbra and a bright penumbra. The penumbra is believed to exhibit magnetoconvection, manifesting itself in the penumbral filaments \citep{2008ApJ...677L.149S,2008A&A...488L..17Z,2009ApJ...691..640R,2011ApJ...729....5R,2014ApJ...785...90R}. The penumbral filaments have a close to horizontal magnetic field, which is interlaced with a more vertical one in the spines \citep{1993ApJ...403..780T,1993A&A...275..283S,1993ApJ...418..928L,2013A&A...557A..25T}. Magnetoconvection is needed in the umbra, as well. Radiative transfer is on its own not sufficient for explaining the observed brightness of the umbra \citep{1965ApJ...141..548D}. The direct manifestation of magnetoconvection in the umbra are the umbral dots \citep{1964ApJ...139...45D,1979ApJ...234..333P,1986ApJ...302..809C,2006ApJ...641L..73S,2010A&A...510A..12B}. Umbral dots are small bright features appearing in umbrae and pores. Depending on their location in the umbra, they are classified as central umbral dots (CUDs), located in the central parts of the umbra, or peripheral umbral dots (PUDs), located in the outer parts of the umbra \citep{1986A&A...156..347G}.

Peripheral umbral dots exhibit many similarities with the heads of penumbral filaments (also referred to as penumbral grains, PGs). Like the heads of the penumbral filaments, many PUDs have an elongated shape \citep{2008ApJ...672..684R,2009A&A...504..575S} and they show upflows and an inclined magnetic field \citep{2009ApJ...694.1080S}. In addition, inward-migrating PGs in some cases turn into PUDs when reaching the umbra-penumbra (UP) boundary \citep{1973SoPh...29...55M,1982SoPh...75...63T}. Simulations also indicate that there are similarities between umbral dots and penumbral filaments \citep{2009ApJ...691..640R}.

The similarities between umbral dots and PGs raise the question why there is a rather sharp UP boundary in sunspots. A possible explanation is that despite these apparent similarities between PUDs and PGs, there still is a discontinuity in the properties of convective features at the UP boundary. Alternatively, the UP boundary could be related to a sudden increase in the number density of convective features. \citet{2018A&A...611L...4J} suggested that the UP boundary occurs where the strength of the vertical magnetic field falls below a fixed threshold. However, this criterion (often called the Jur{\v c}{\'a}k criterion) does not take into account that the penumbra consists of both filaments and spines, whose properties vary between spots of different sizes \citep{2020A&A...639A.106L}.

Thus, a first step towards understanding the origin of the UP boundary is a detailed characterization of the convective elements in the umbra and in the penumbra and how they change across the UP boundary. Here we used observations of two sunspots (one rather small and one large spot) made with Hinode to compare the properties of umbral dots and penumbral filaments and to determine how these change between spots of different sizes. We derived ensemble averages of different types of features following the approach of \citet{2013A&A...557A..25T}.

\section{Data and methods}

\subsection{Hinode observations} \label{sect:data}
\begin{figure*}
\includegraphics[width=\linewidth]{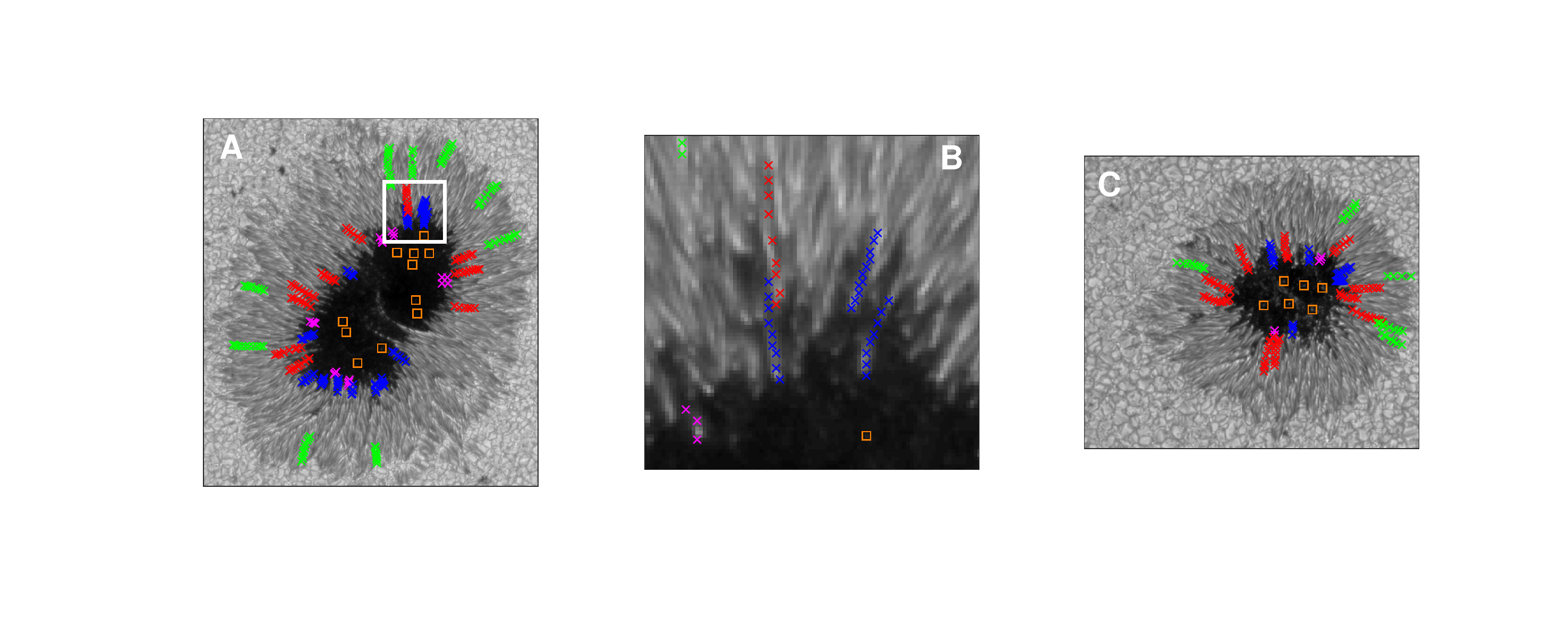}
\caption{Maps of the continuum intensity obtained by Hinode/SOT of the two sunspots used in this study. {\it Panel A:} the large sunspot AR~10923 observed on 14 November 2006, {\it panel B:} zoom of a region around the umbra-penumbra boundary of AR~10923 (as indicated by the white rectangle in panel A), and {\it panel C:} the small sunspot AR~10933 observed on 6 January 2007. The symbols indicate the individual features used in this study. Orange: CUDs, purple: short PUDs, blue: long PUDs, red: penumbral filaments in the inner penumbra, and green: penumbral filaments in the outer penumbra. See text for more details on how we selected the individual features.}
\label{fig:spots}
\end{figure*}

We derived the properties of umbral dots and penumbral filaments for two sunspots, a rather large spot (AR~10923 observed on 14 November 2006) and a smaller one (AR~10933 observed on 6 January 2007). These were observed with the spectropolarimeter on the Solar Optical Telescope \citep[SOT/SP,][]{2007SoPh..243....3K,2008SoPh..249..167T,2008SoPh..249..233I,2013SoPh..283..579L} onboard the Hinode spacecraft. This instrument performs spectropolarimetric observations using the Fe~I line pair at $6301.5$~\AA \ and $6302.5$~\AA. The sunspots were observed in normal mode, with a spatial sampling of $0.16''$ per pixel. Figure~\ref{fig:spots} shows continuum intensity images of the two sunspots. Both sunspots were located close to disk center at the time of the observation. The heliocentric angle $\theta$ was $8^\circ$ for AR~10923 and $9^\circ$ for AR~10933. Hence, the appearance of the umbral dots and penumbral filaments is not or at the most very slightly affected by projection or radiative transfer effects that would arise from an inclined line-of-sight. The spots exhibited regular shapes with only one umbral core each and fully-fledged penumbrae. The areas of the spots were 2698~Mm$^2$ (for the spot in AR~10923) and 1003~Mm$^2$ (for the spot in AR~10933) at the time of the observations. This difference in size allows to infer how the properties of umbral dots and penumbral filaments depend on the size of the host sunspot. Also, both spots were stable at the time of the observations (the shape and size of both sunspots did not change considerably over the course of a few days).

We inverted the Stokes parameters observed by Hinode for both sunspots in order to derive the height-dependent atmospheric parameters. We used the spatially coupled version of the SPINOR code \citep{2000A&A...358.1109F,2012A&A...548A...5V,2013A&A...557A..24V}, which assumes local thermodynamic equilibrium (LTE). We set three nodes in optical depth, placed at $\log{\tau} = -2.5,-0.9,0$, \citep[cf. ][]{2013A&A...557A..25T}. We calibrated the line-of-sight velocity by defining $v_{\rm los} = 0$ as the mean velocity of the umbra while ignoring outliers (pixels, where $|v_{\rm los}| > 10^3$~m/s at $\log \tau = -0.9$). The outliers in the inverted velocity are caused by the presence of molecular lines in the dark parts of the umbra. The inversion becomes unreliable for pixels with a strong signature of molecular lines. In most cases, the maps of the inverted velocity exhibit extremely large amplitudes at these locations (mostly downflows of a few km/s).

We then resolved the $180^\circ$ azimuthal ambiguity by using the non-potential magnetic field computation method \citep[NPFC,][]{2005ApJ...629L..69G}  and  transformed the magnetic field vector to the local reference frame.

\subsection{Selection of features and ensemble averaging} \label{sect:features}

\begin{figure*}
\includegraphics[width=\linewidth]{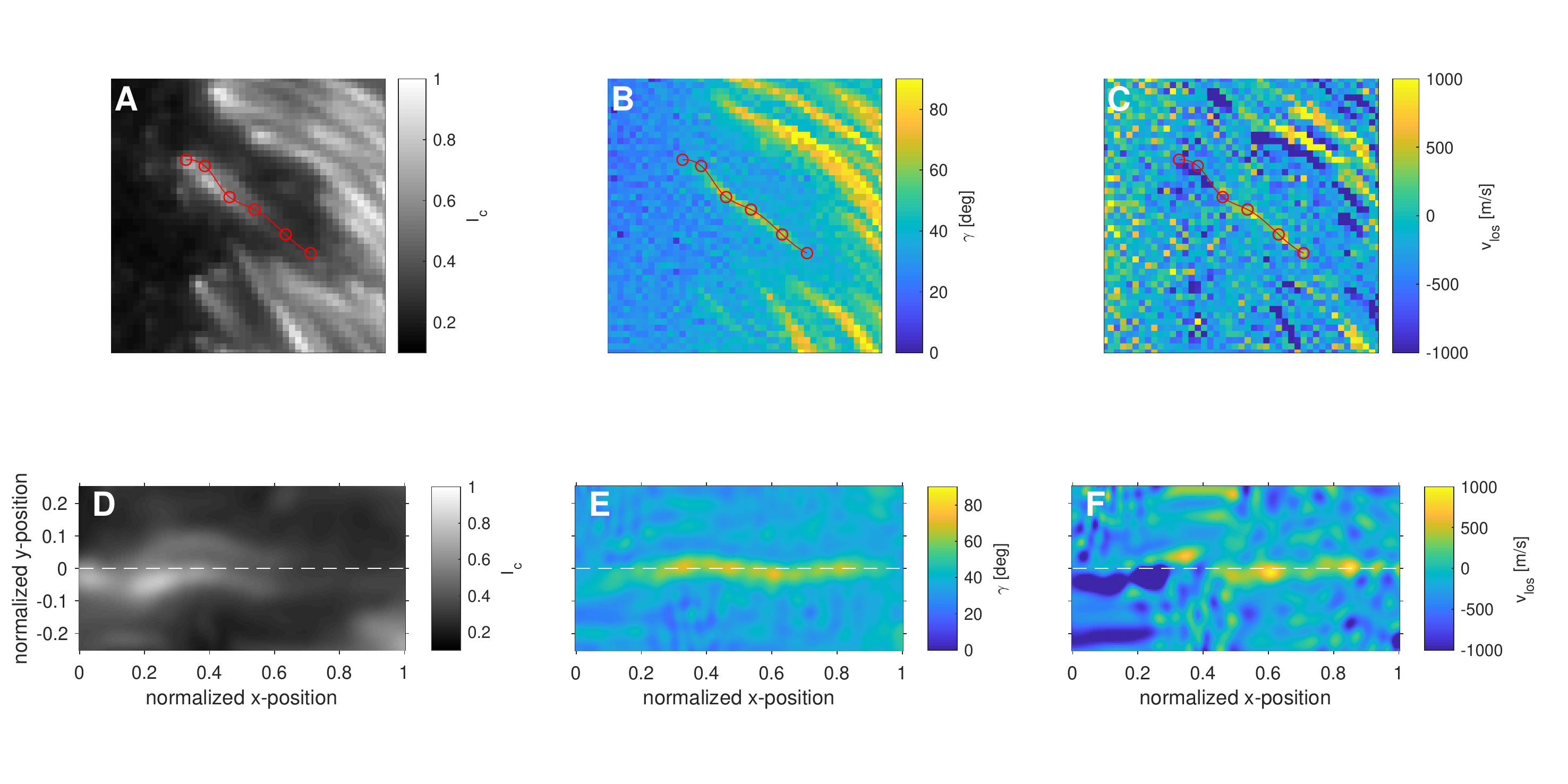}
\caption{Selection of individual peripheral umbral dots. We identified individual PUDs using maps of the continuum intensity ({\it panel~A}), of the inclination of the magnetic field vector with respect to the local vertical axis ({\it panel~B}), and of the line-of-sight velocity ({\it panel~C}). The last two observables are derived from the inversion at $\tau = 1$. We manually selected pixels along the main axis of the PUD (indicated by the red circles in panels~A to~C). These individual points are then connected using a spline interpolation (solid red curves in panels~A to~C). This curve is then used to retrieve the de-stretched, de-rotated, and length-normalized PUDs (shown in {\it panels~D to~F}). See Sect.~\ref{sect:features} for more details on the identification of individual features. This procedure is analogous to the identification of individual penumbral filaments in \citet{2013A&A...557A..25T}.}
\label{fig:PUD_selection}
\end{figure*}

Our aim is to compare the properties of convective features of the two sunspots by deriving ensemble averages of the various types of convective elements in sunspots. These ensemble averages bring out the common properties of the selected elements \cite[]{2013A&A...557A..25T}. Here we distinguished between penumbral filaments (located either in the inner or in the outer penumbra), PUDs, and CUDs. We separated between CUDs and PUDs by using the location of the features and their shape. The CUDs are located in the inner part of the umbra and exhibit a roundish shape. The PUDs are located closer to the UP boundary and are more elongated than CUDs. Some of the PUDs are not fully resolved by Hinode/SOT/SP data and therefore their shape cannot be determined. We only selected PUDs that are spatially resolved and that exhibit an aspect ratio (length of the PUD divided by its width) of at least two. We also divided the PUDs into two groups. Some PUDs exhibit downflows at $\tau = 1$ in their tails, while others do not show clear evidence for the existence of downflows. We found that the PUDs with downflows are more elongated and are located closer to the UP boundary than their counterparts without evidence of downflows. Most likely, however, all of the PUDs exhibit downflows, but in some cases they cannot be resolved in our data. This could be either due to the limited spatial resolution of Hinode or due to the presence of noise in the velocity maps. Without a clear signature of a downflow, we cannot trace the tails of the PUDs. Since we do not want to combine features with and without tails in the ensemble average, we distinguished between these two types of PUDs (short and long PUDs). Table~\ref{tab:feats} lists the different types of convective features and our criteria for distinguishing between them. For each type of convective element, we then identified several individual features, both in AR~10923 and in AR~10933. Unfortunately, there is no easy automatic method yet for identifying individual penumbral filaments. Therefore, we applied a manual procedure, following \citet{2013A&A...557A..25T}. 

Identifying individual penumbral filaments is easiest when considering the continuum intensity and the maps of the inclination of the magnetic field as well as maps of the line-of-sight velocity (both at $\tau = 1$) returned by the inversion. The head of a penumbral filament is bright and exhibits a strong upflow. It is followed by the body of the penumbral filament, which is most discernible by its highly inclined magnetic field. The tails of the penumbral filaments can then be identified by their strong downflows. We manually identified a few points along the central axis of the features. Then, a line connecting these points was computed using a spline interpolation, from which 201 equidistantly distributed points were selected. In combination with a grid of 51~points perpendicular to the central axis of the feature, this defines a new grid, upon which we interpolated the maps of the continuum intensity ($I_{\rm C}$), the vertical magnetic field ($B_{\rm z}$), the inclination of the magnetic field to the local vertical ($\gamma$), and the line-of-sight velocity ($v_{\rm los}$), all at $\tau = 1$. This resulted in de-stretched, de-rotated, and length-normalized maps for each individual penumbral filament.

We identified individual PUDs and extracted their atmospheric parameters in a similar way. We traced them by using continuum intensity and the inclination of the magnetic field. The main difference to penumbral filaments is that we do not observe a reversal of the magnetic polarity at the tails of PUDs compared to the polarity at their heads. As explained above, we classify the PUDs as short or long PUDs depending on their length and whether we could trace the tail of the PUDs in the velocity maps. Since PUDs are less elongated than penumbral filaments, we resample them on a grid of $201 \times 101$~grid points (for long PUDs) or $201 \times 151$~grid points (for short PUDs). Figure~\ref{fig:PUD_selection} visualizes the method for identifying individual PUDs. In case of the CUDs, we extracted a region of $11\times 11$~pixels around the center of the umbral dot, which was identified by hand. Since the CUDs have a roundish shape, they do not need to be de-stretched or de-rotated. Hence, we do not need to interpolate them on a new grid.

The number of features that were identified in this way is given in Table~\ref{tab:feats}. Their location in the sunspots is indicated by the crosses in Figure~\ref{fig:spots}, with the colours identifying the types of features.

\begin{table*}
\caption{Overview of the criteria used to identify the different types of features studied here.}
\label{tab:feats}
\centering
\begin{tabular}{l l l l l l l}
\hline\hline
Type of feature & elongated & downflows & polarity reversal & \# in AR~10923 & \# in AR~10933 \\
\hline
Penumbr. fil. (outer penumbra) & Yes & Yes & Yes &  9 &  5 \\ 
Penumbr. fil. (inner penumbra) & Yes & Yes & Yes & 10 & 10 \\ 
long PUD & Yes & Yes & No & 12 & 6 \\
short PUD & Yes & Unclear & No & 7 & 2 \\
CUD & No & No & No & 11 & 6 \\
\hline
\end{tabular}
\end{table*}

\section{Results}

\subsection{Ensemble averages}
\begin{figure*}
\includegraphics[width=\linewidth]{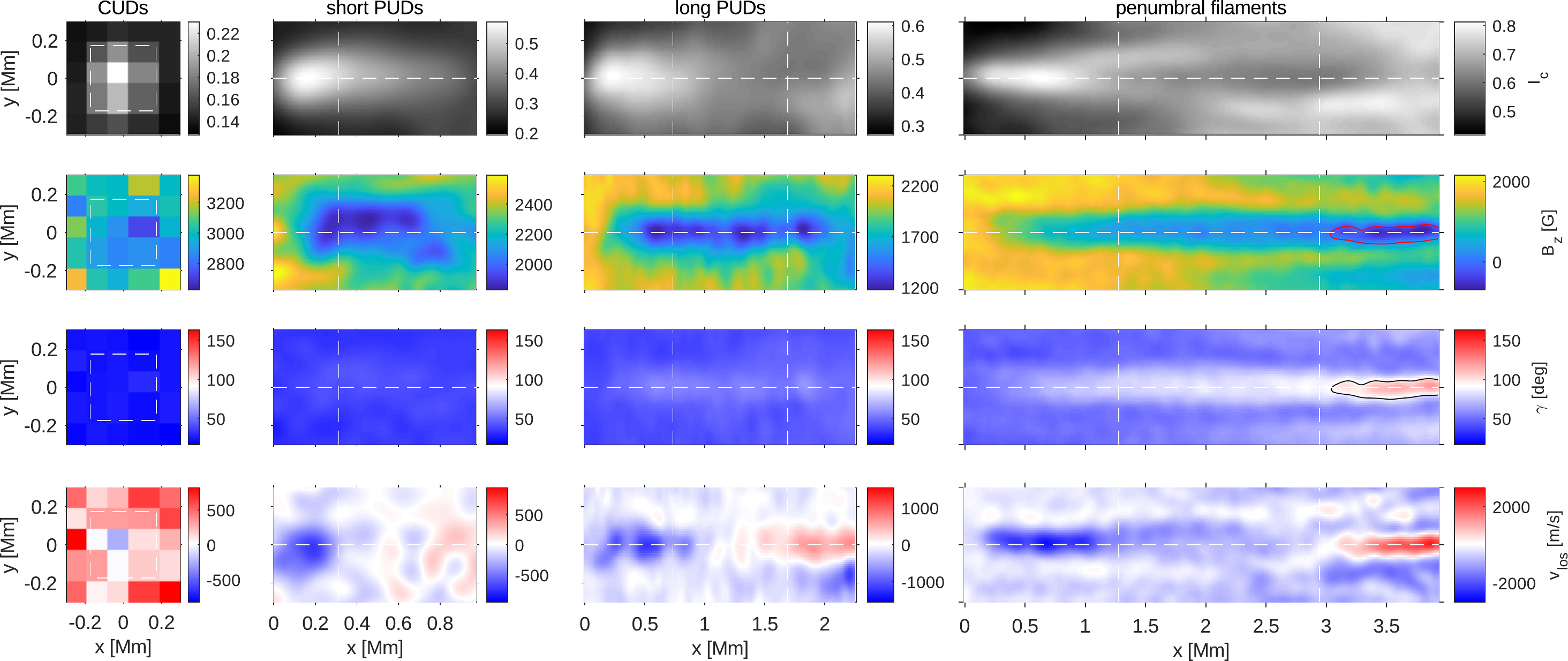}
\caption{Ensemble averages of various observables for the different types of features in AR~10923 (see Figure~\ref{fig:spots}). {\it From left to right:} CUDs, short PUDs, long PUDs, and penumbral filaments in the inner penumbra. {\it From top to bottom:} continuum intensity, vertical magnetic field, inclination of the magnetic field, and line-of-sight velocity. The white vertical lines separate the heads, bodies, and tails of the PUDs and of the penumbral filaments. In case of the short PUDs, we only separate between the head and the body, since we do not observe tails for these features when considering them individually. The dashed squares in the left column outline the umbral dot and the horizontal lines in the other columns indicate the central axis of the features. The red contours in the maps of $B_{\rm z}$ and the black contours in the maps of $\gamma$ indicate regions where the polarity of the field changes. The range of the $y$-axis is $\pm 0.3$~Mm in all panels except for the ones showing the results for the penumbral filaments. Here, the range of the $y$-axis is $\pm 0.5$~Mm. We note that the aspect ratios of the plots in columns 2, 3 and 4 are not correct.}
\label{fig:ens_avg}
\end{figure*}

\begin{figure*}
\includegraphics[width=\linewidth]{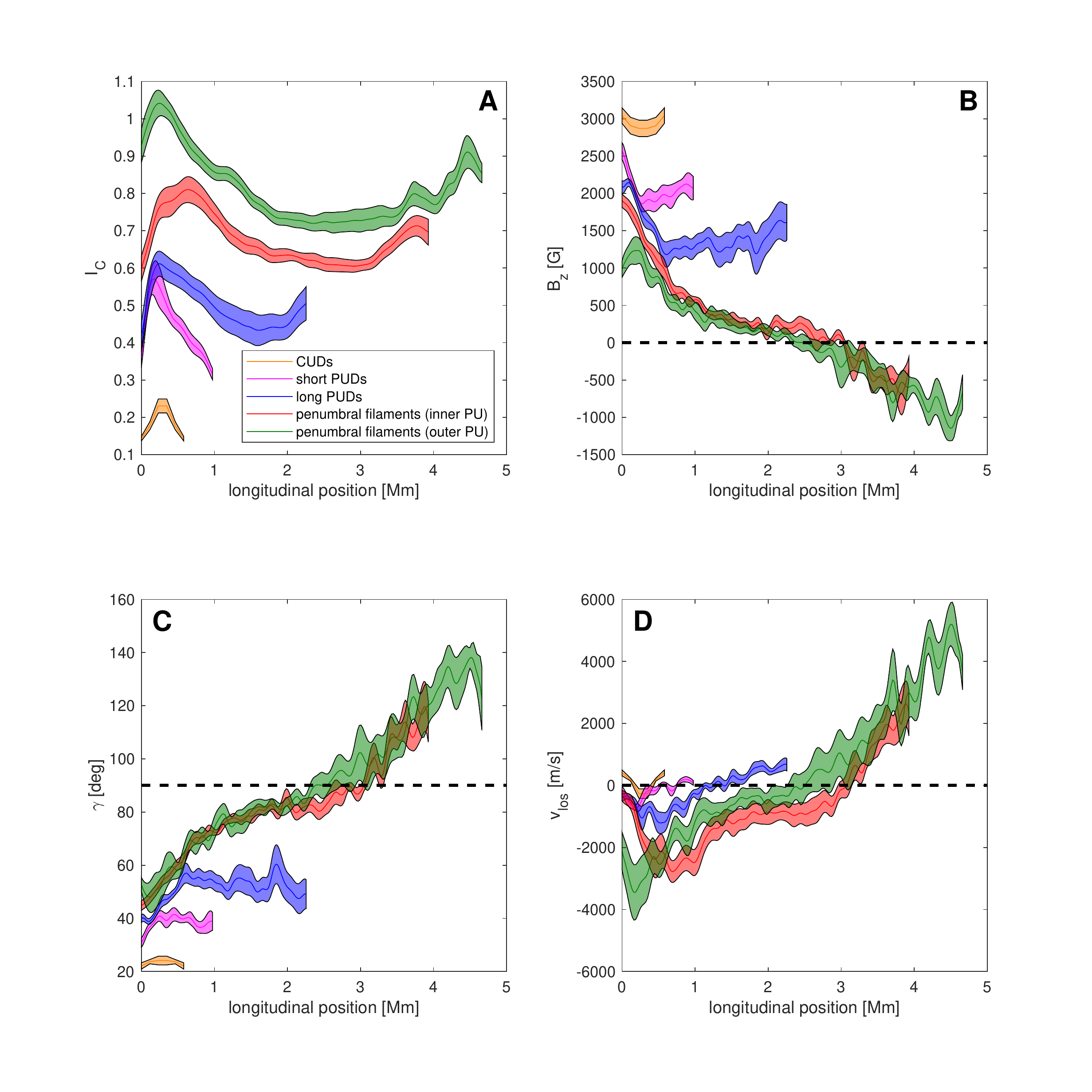}
\caption{Cuts along the central axis of the ensemble averages of the different types of features in AR~10923 shown in Figure~\ref{fig:ens_avg} for various observables. For the CUDs, we performed an azimuthal average. {\it Panel~A}: continuum intensity, {\it panel~B}: vertical magnetic field, {\it panel~C}: inclination of the magnetic field, and {\it panel~D}: line-of-sight velocity. Orange: CUDs, purple: short PUDs, blue: long PUDs, red: penumbral filaments in the inner penumbra, and green: penumbral filaments in the outer penumbra. The shaded areas indicate the $1\sigma$ error of the mean.}
\label{fig:avg_cuts}
\end{figure*}

Since we identified several individual features for each type of convective element, we can now derive ensemble averages. Figure~\ref{fig:ens_avg} shows maps of the ensemble averages for various observables for the different types of features discussed in Section~\ref{sect:features} and Figure~\ref{fig:avg_cuts} shows cuts along the central axis of the ensemble averages (an azimuthal average in case of the CUDs). We only show the results for the large sunspot AR~10923, since the results for AR~10933 look qualitatively similar. Generally, when going from the outer penumbra to the center of the umbra, convective features become shorter, darker, exhibit stronger and more vertical magnetic field and have weaker upflows.

The resulting average penumbral filaments are largely consistent with the results of \citet{2013A&A...557A..25T}. Penumbral filaments can be divided into three different parts, a head, a body, and a tail. The heads of penumbral filaments are bright, have a relatively strong vertical magnetic field with the same polarity as the umbra, and also exhibit an upflow. The heads are followed by a region with a highly inclined magnetic field, that is similar in brightness to the surroundings. The tails of the penumbral filaments are bright again (although less bright than the heads). They have strong vertical magnetic fields with the polarity being opposite to the one of the umbra and they show strong downflows. Penumbral filaments also exhibit lateral downflows along their sides \citep{2011Sci...333..316S,2011ApJ...734L..18J,2012A&A...540A..19S,2013A&A...557A..25T,2015ApJ...803...93E}. While our map of $v_{\rm los}$ of the average penumbral filament shows the presence of two lateral bands located on both sides of the penumbral filament, the line-of-sight velocity that we measure in these bands is actually negative (indicating a weak upflow of about $100\-- 200$~m/s). This is in spite of the fact that we detected lateral downflows within individual penumbral filaments \citep[with downflow velocities of a few 100~m/s, in agreement with][]{2013A&A...557A..25T}. These lateral downflow lanes are very narrow and the velocity of the downflows is much weaker than the velocity of the upflow along the main axis of the penumbral filaments ($1\-- 2$~km at the center of the penumbral filaments). Since the individual penumbral filaments have slightly different shapes and widths, the weak signal of the lateral downflows gets mixed with the stronger upflow along the main axis when deriving the ensemble average. In addition, the upflow along the main axis of the penumbral filaments is enhanced by leakage from the Evershed flow. AR~10923 was not observed exactly at disk center ($\theta = 8^\circ$) and most of the penumbral filaments that we selected in AR~10923 are located on the diskward side of the penumbra, where the Evershed flow appears as a blueshift. This causes the lateral bands to exhibit an apparent upflow in the ensemble average. We note that \cite{2017A&A...607A..36S} did not observe lateral downflows in their ensemble average of penumbral filaments for a sunspot that was observed away from disk center, either. The bias in the velocity map of the averaged penumbral filament does not severely affect the subsequent analysis, though. The lateral downflows contribute much less to the returning downflow in penumbral filaments than the flows in their tails \citep{2013A&A...557A..25T} in terms of mass conservation. Therefore, we focus in this study on flows along the main axis of the penumbral filament.

The penumbral filaments in the inner and in the outer penumbra look very similar, although the penumbral filaments in the outer penumbra are a bit longer and brighter than the ones in the inner penumbra. Deduced differences between filaments in the inner and outer penumbra may be affected by stray light. The presence of stray light causes features in the vicinity of the umbra to appear darker. Similarly, the brightness of penumbral filaments in the outer penumbra is slightly enhanced by stray light from the bright granulation outside the sunspot. However, the level of stray light in Hinode SOT/SP is very low \citep[about 1\% at 10'' distance,][]{2013SoPh..283..579L} and so, we do not expect it to be the main reason why we observe the penumbral filaments in the outer penumbra to be brighter than the ones in the inner penumbra.

The PUDs look in many aspects similar to penumbral filaments. They also have an elongated shape with a bright head. However, they are much shorter than penumbral filaments and not as bright. In addition, the magnetic field of PUDs is stronger and more vertical than the one of penumbral filaments. When performing an ensemble average, even the short PUDs exhibit weak downflows (up to $\sim 230$~m/s) in their tails. This downflow is too weak to be detected for individual PUDs.
The most prominent difference between PUDs and penumbral filaments is that we do not observe a polarity reversal at the tails of PUDs. Similarities between PUDs and PGs have already been reported by \citet{2009ApJ...694.1080S}.

The CUDs are darker than the PUDs and have even stronger and more vertical magnetic fields. Unfortunately, many umbral dots are not fully resolved in the Hinode data. Previous measurements of the diameter of umbral dots range from 180 to 300~km \citep[e. g.,][]{1997A&A...328..682S,2007PASJ...59S.585K,2012ApJ...752..109L}, with the smaller ones lying below the resolution limit of Hinode/SOT/SP.

The mass motions within umbral dots are not well understood yet. In numerical simulations, umbral dots have a central upflow, with surrounding downflows \citep{2006ApJ...641L..73S}. However, there are conflicting results from observations, particularly on the (non)-existence of downflows \citep[e. g.,][]{2004ApJ...614..448S,2007ApJ...665L..79B,2008ApJ...678L.157R,2013A&A...554A..53R,2009ApJ...694.1080S,2009ApJ...702.1048W,2012ApJ...757...49W,2010ApJ...713.1282O}. Here, we can detect up- and downflows in all types of convective features. For the CUDs, it is unclear if the downflow signal in Fig.~\ref{fig:ens_avg} represents an actual flow. As explained in Sect.~\ref{sect:data}, the presence of molecular lines can lead to apparent strong downflows in the umbra.

\subsection{Dependence on the length of the features}
As shown in the previous section, the physical properties of the convective features change smoothly from CUDs to penumbral filaments. In addition, the length of the features also increases from the center of the umbra to the outer penumbra. This suggests that there might be a connection between the length and the physical parameters of the features.

\begin{figure*}
\includegraphics[width=\linewidth]{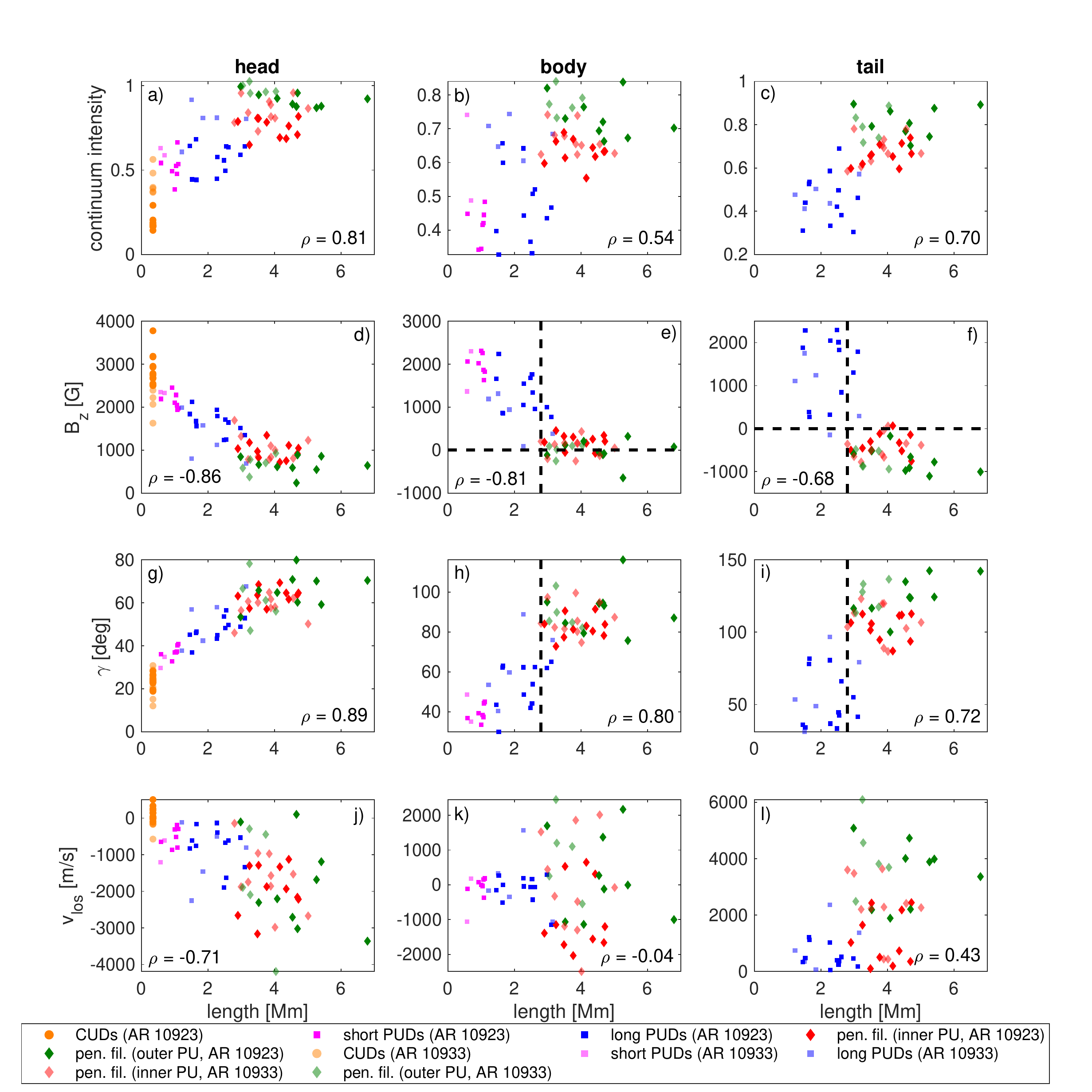}
\caption{Scatter plots of various observables as a function of the length of the individual features for both sunspots. Each data point represents one feature and is an average along the main axis over the head ({\it left column}), the body ({\it center column}), or the tail ({\it right column}) of the feature. These regions are separated by the white vertical lines in Figure~\ref{fig:ens_avg}. {\it From top to bottom:} continuum intensity, vertical magnetic field, inclination of the magnetic field, and line-of-sight velocity. Orange: CUDs, purple: short PUDs, blue: long PUDs, red: penumbral filaments in the inner penumbra, and green: penumbral filaments in the outer penumbra. The solid symbols show the results for AR~10923 and the transparent ones for AR~10933. We also show in each panel the correlation coefficient $\rho$ between the respective physical quantity and the length of the features. We do not distinguish between the different types of features when computing the correlation coefficient. Since the short PUDs do not exhibit tails, we only separate between the head and the body of these features. The CUDs resemble the heads of the penumbral filaments, so we include these features only in the left column. We average the CUDs in a region of $3\times 3$~pixels around their centers (as indicated by the dashed white squares in the left column of Figure~\ref{fig:ens_avg}). Since many of the CUDs are not fully resolved in the Hinode data, we assigned a length of 3~pixels ($\sim 350$~km) to each of them.} 
\label{fig:length}
\end{figure*}

As can be seen in Figure~\ref{fig:length}, all observables (except for maybe the velocity) are monotonic functions of the length. In addition, there is a smooth transition of the physical parameters between the different types of convective elements. However, as discussed above, the polarity of the magnetic field reverses in the tails of penumbral filaments, but not in the other types of convective features. This causes a strong change of the inclination and of $B_{\rm z}$ at a length of about 2.8~Mm (as indicated by the vertical lines in some of the panels in Figure~\ref{fig:length}). Apart from that, there is a smooth transition from umbral dots to penumbral filaments. The properties of the features are solely determined by their length. The dependence on length is the strongest for the heads of the features (see the correlation coefficients in Figure~\ref{fig:length}).

There are no differences in the ensemble averages for AR~10923 and AR~10933, apart from an offset in the continuum intensity. In the smaller spot, the features exhibit slightly higher continuum intensities in the head and in the body than in the larger spot.

\section{Discussion}
We observed similarities in the properties of different types of convective features in sunspots. There is a smooth transition from umbral dots to penumbral filaments with most physical parameters of the features being continuous functions of the length of the features. Also, apart from an offset in the continuum intensity, there are no obvious differences in the properties of the convective elements between the large and small spot that we analyzed. The penumbral filaments and umbral dots in the small spot are brighter than the ones in the large spot, even though their size is comparable. This decrease of the brightness of penumbral filaments was already reported by \citet{2020A&A...639A.106L}.

The most striking difference between PUDs and penumbral filaments is that only the latter exhibit a polarity reversal in their tails relative to the polarity of the umbra. The absence of a polarity reversal in our observations of the tails of PUDs is unexpected, since these features exhibit overturning convection. Most likely, there is a polarity reversal, but it cannot be detected by Hinode. Since the downflow velocities in the tails of the PUDs are very weak (much weaker than for the penumbral filaments), the field strength of the opposite polarity field in the tails should be quite low. Hence, it could easily be obscured by the magnetic field of the surroundings or of higher atmospheric layers. Alternatively, the polarity reversal could occur below the visible surface. The tails of PUDs at $\tau = 1$ look similar to the tails of penumbral filaments at $\log \tau = -2.5$. Similar to PUDs at $\tau = 1$, the tails of short penumbral filaments exhibit only weak downflows ($\sim 200$ms$^{-1}$) and inclinations of about $60^\circ$ at $\log \tau = -2.5$ \citep{2013A&A...557A..25T}.

These similarities between PUDs and penumbral filaments pose the question why there is such a rapid increase in brightness at the UP boundary. There are two possible explanations for this. The magnetoconvection in penumbral filaments could be much more effective than the one in PUDs. Alternatively, the number density of convective features could be higher in the penumbra than in the umbra. 

We derived estimates of the number densities of PUDs and penumbral filaments by manually identifying features in a small part of the sunspot in AR~10923. In case of the PUDs, we focused only on bright, elongated PUDs in the outermost umbra. We counted the PUDs in a small patch of the outer umbra using an image of the continuum intensity. For the penumbral filaments, we inferred the number of filament tails in an image of the magnetic inclination at $\tau = 1$ for a small patch of the inner penumbra. The resulting number densities are almost identical ($\sim 0.35$~1/Mm$^2$ for the PUDs and $\sim 0.36$~1/Mm$^2$ for the penumbral filaments.) Our estimate of the number density of PUDs is consistent with the results of \citet{2014PASJ...66S...1W} for UDs ($0.1 \-- 0.43$~1/Mm$^2$), but it is lower than the one of \citet{2012ApJ...745..163K} ($1.9$~1/Mm$^2$). Thus, our results suggest that there are no significant differences in the number density of convective features between the outermost part of the umbra and the inner penumbra.

Numerical simulations suggest that the vigorous magnetoconvection in penumbral filaments is very efficient in transporting energy and that it is sufficient for explaining the brightness of the penumbra \citep[e.g.,][]{2007ApJ...669.1390H,2008ApJ...677L.149S,2009ApJ...691..640R}. Our results also indicate that penumbral filaments are much brighter, more extended, and harbour much stronger flow velocities than the PUDs in the outer parts of the umbra. We cannot rule out, however, that a part of the excess brightness, velocities, etc. of penumbral filaments relative to UDs is due to the fact that they are better resolved by the observations. We do not expect, however, that this on its own explains the difference. Since penumbral filaments are significantly larger than PUDs, their area filling factor is larger, as well. The combination of the higher brightness (especially at their heads, which border the umbra) with the larger filling factor of the penumbral filaments explains the rapid increase in brightness at the UP boundary.

Fig.~\ref{fig:length} suggests that the length and the brightness of convective features increases with decreasing strength of the vertical magnetic field and increasing inclination. Unfortunately, our results do not allow us to discriminate between these two factors (the correlation coefficients are similar, as can be deduced directly from the figure). We also note that we show in Fig.~\ref{fig:length} the $B_{\rm z}$ and $\gamma$ of the features themselves. However, the convection alters the magnetic field within the convective cells. Therefore, the magnetic field within the convective features does not represent the conditions in which the convective cells formed. The larger size and higher brightness of penumbral filaments compared to UDs are probably caused by differences in the strength and geometry of the large-scale magnetic field in the umbra and in the penumbra. While the magnetic field is strong and almost vertical in the umbra, it is weaker and more inclined in the penumbra.

\citet{2018A&A...611L...4J} observed that the mean strength of $B_{\rm z}$ at the UP boundary is constant in their sample of spots. They interpreted this constant value of $B_{\rm z}$ to be the threshold for the transition from umbra to penumbra. However, their interpretation was later challenged by \citet{2020A&A...639A.106L}, who attributed this constant $B_{\rm z}$ to be caused by the decrease of the brightness of penumbral filaments with increasing spot size (see also Fig.~\ref{fig:length}). In addition, penumbral filaments are elongated structures that are aligned with the direction of the horizontal component of the magnetic field in the penumbra. This alignment indicates that a sufficiently large inclination of the magnetic field is also important for the formation of penumbral filaments and so, a low $B_{\rm z}$ on its own is not sufficient for explaining the formation of the penumbra. Indeed, numerical simulations suggest that the elongation of convective cells is governed by the inclination of the underlying magnetic field in the subsurface layers \citep[e.g.,][]{2009ApJ...691..640R}. The higher the inclination, the more elongated are the convective features. However, unlike $B_{\rm z}$, the inclination does not have a fixed value at the UP boundary, but increases with increasing spot size \citep{2018A&A...611L...4J}. This suggests that there is no threshold of the inclination that would trigger the formation of the penumbra. Most likely, the size, shape, and brightness of convective features are affected by both, the strength and the inclination of the magnetic field. Currently, it is unclear, which combinations of $B_{\rm z}$ and $\gamma$ can give rise to penumbral filaments. Disentangling the influence of these two factors is further impeded by the fact that they are not independent from each other in sunspots.

Both, the inclination, and $B_{\rm z}$ do not exhibit a discontinuity at the UP boundary, but increase smoothly. Nevertheless, the change of these parameters can be responsible for the rapid increase of the brightness at the UP boundary. As discussed above, both the size and the brightness of convective features depend strongly on the properties of the underlying magnetic field. The combination of these two factors causes the integrated brightness of convective cells to be very sensitive to changes of the magnetic field strength and geometry.


\begin{acknowledgements}
This work benefited from the Hinode sunspot database at MPS, created by Gautam Narayan. This project has received funding from the European Research Council (ERC) under the European Union’s Horizon 2020 research and innovation programme (grant agreement No 695075) and has been supported by the BK21 plus program through the National Research Foundation (NRF) funded by the Ministry of Education of Korea. Hinode is a Japanese mission developed and launched by ISAS/JAXA, collaborating with NAOJ as a domestic partner, NASA and STFC (UK) as international partners. Scientific operation of the Hinode mission is conducted by the Hinode science team organized at ISAS/JAXA. This team mainly consists of scientists from institutes in the partner countries. Support for the post-launch operation is provided by JAXA and NAOJ (Japan), STFC (U.K.), NASA, ESA, and NSC (Norway).
\end{acknowledgements}

\bibliographystyle{aa} 
\bibliography{literature} 

\begin{thebibliography}{49}
\expandafter\ifx\csname natexlab\endcsname\relax\def\natexlab#1{#1}\fi

\bibitem[{{Bharti} {et~al.}(2010){Bharti}, {Beeck}, \&
  {Sch{\"u}ssler}}]{2010A&A...510A..12B}
{Bharti}, L., {Beeck}, B., \& {Sch{\"u}ssler}, M. 2010, \aap, 510, A12

\bibitem[{{Bharti} {et~al.}(2007){Bharti}, {Jain}, \&
  {Jaaffrey}}]{2007ApJ...665L..79B}
{Bharti}, L., {Jain}, R., \& {Jaaffrey}, S.~N.~A. 2007, \apjl, 665, L79

\bibitem[{{Choudhuri}(1986)}]{1986ApJ...302..809C}
{Choudhuri}, A.~R. 1986, \apj, 302, 809

\bibitem[{{Danielson}(1964)}]{1964ApJ...139...45D}
{Danielson}, R.~E. 1964, \apj, 139, 45

\bibitem[{{Deinzer}(1965)}]{1965ApJ...141..548D}
{Deinzer}, W. 1965, \apj, 141, 548

\bibitem[{{Esteban Pozuelo} {et~al.}(2015){Esteban Pozuelo}, {Bellot Rubio}, \&
  {de la Cruz Rodr{\'\i}guez}}]{2015ApJ...803...93E}
{Esteban Pozuelo}, S., {Bellot Rubio}, L.~R., \& {de la Cruz Rodr{\'\i}guez},
  J. 2015, \apj, 803, 93

\bibitem[{{Frutiger} {et~al.}(2000){Frutiger}, {Solanki}, {Fligge}, \&
  {Bruls}}]{2000A&A...358.1109F}
{Frutiger}, C., {Solanki}, S.~K., {Fligge}, M., \& {Bruls}, J.~H.~M.~J. 2000,
  \aap, 358, 1109

\bibitem[{{Georgoulis}(2005)}]{2005ApJ...629L..69G}
{Georgoulis}, M.~K. 2005, \apjl, 629, L69

\bibitem[{{Grossmann-Doerth} {et~al.}(1986){Grossmann-Doerth}, {Schmidt}, \&
  {Schroeter}}]{1986A&A...156..347G}
{Grossmann-Doerth}, U., {Schmidt}, W., \& {Schroeter}, E.~H. 1986, \aap, 156,
  347

\bibitem[{{Heinemann} {et~al.}(2007){Heinemann}, {Nordlund}, {Scharmer}, \&
  {Spruit}}]{2007ApJ...669.1390H}
{Heinemann}, T., {Nordlund}, {\r{A}}., {Scharmer}, G.~B., \& {Spruit}, H.~C.
  2007, \apj, 669, 1390

\bibitem[{{Ichimoto} {et~al.}(2008){Ichimoto}, {Lites}, {Elmore}, {Suematsu},
  {Tsuneta}, {Katsukawa}, {Shimizu}, {Shine}, {Tarbell}, {Title}, {Kiyohara},
  {Shinoda}, {Card}, {Lecinski}, {Streander}, {Nakagiri}, {Miyashita},
  {Noguchi}, {Hoffmann}, \& {Cruz}}]{2008SoPh..249..233I}
{Ichimoto}, K., {Lites}, B., {Elmore}, D., {et~al.} 2008, \solphys, 249, 233

\bibitem[{{Joshi} {et~al.}(2011){Joshi}, {Pietarila}, {Hirzberger}, {Solanki},
  {Aznar Cuadrado}, \& {Merenda}}]{2011ApJ...734L..18J}
{Joshi}, J., {Pietarila}, A., {Hirzberger}, J., {et~al.} 2011, \apjl, 734, L18

\bibitem[{{Jur{\v c}{\'a}k} {et~al.}(2018){Jur{\v c}{\'a}k}, {Rezaei},
  {Gonz{\'a}lez}, {Schlichenmaier}, \& {Vomlel}}]{2018A&A...611L...4J}
{Jur{\v c}{\'a}k}, J., {Rezaei}, R., {Gonz{\'a}lez}, N.~B., {Schlichenmaier},
  R., \& {Vomlel}, J. 2018, \aap, 611, L4

\bibitem[{{Kilcik} {et~al.}(2012){Kilcik}, {Yurchyshyn}, {Rempel}, {Abramenko},
  {Kitai}, {Goode}, {Cao}, \& {Watanabe}}]{2012ApJ...745..163K}
{Kilcik}, A., {Yurchyshyn}, V.~B., {Rempel}, M., {et~al.} 2012, \apj, 745, 163

\bibitem[{{Kitai} {et~al.}(2007){Kitai}, {Watanabe}, {Nakamura}, {Otsuji},
  {Matsumoto}, {UeNo}, {Nagata}, {Shibata}, {Muller}, {Ichimoto}, {Tsuneta},
  {Suematsu}, {Katsukawa}, {Shimizu}, {Tarbell}, {Shine}, {Title}, \&
  {Lites}}]{2007PASJ...59S.585K}
{Kitai}, R., {Watanabe}, H., {Nakamura}, T., {et~al.} 2007, \pasj, 59, S585

\bibitem[{{Kosugi} {et~al.}(2007){Kosugi}, {Matsuzaki}, {Sakao}, {Shimizu},
  {Sone}, {Tachikawa}, {Hashimoto}, {Minesugi}, {Ohnishi}, {Yamada}, {Tsuneta},
  {Hara}, {Ichimoto}, {Suematsu}, {Shimojo}, {Watanabe}, {Shimada}, {Davis},
  {Hill}, {Owens}, {Title}, {Culhane}, {Harra}, {Doschek}, \&
  {Golub}}]{2007SoPh..243....3K}
{Kosugi}, T., {Matsuzaki}, K., {Sakao}, T., {et~al.} 2007, \solphys, 243, 3

\bibitem[{{Lites} {et~al.}(2013){Lites}, {Akin}, {Card}, {Cruz}, {Duncan},
  {Edwards}, {Elmore}, {Hoffmann}, {Katsukawa}, {Katz}, {Kubo}, {Ichimoto},
  {Shimizu}, {Shine}, {Streander}, {Suematsu}, {Tarbell}, {Title}, \&
  {Tsuneta}}]{2013SoPh..283..579L}
{Lites}, B.~W., {Akin}, D.~L., {Card}, G., {et~al.} 2013, \solphys, 283, 579

\bibitem[{{Lites} {et~al.}(1993){Lites}, {Elmore}, {Seagraves}, \&
  {Skumanich}}]{1993ApJ...418..928L}
{Lites}, B.~W., {Elmore}, D.~F., {Seagraves}, P., \& {Skumanich}, A.~P. 1993,
  \apj, 418, 928

\bibitem[{{L{\"o}ptien} {et~al.}(2020){L{\"o}ptien}, {Lagg}, {van Noort}, \&
  {Solanki}}]{2020A&A...639A.106L}
{L{\"o}ptien}, B., {Lagg}, A., {van Noort}, M., \& {Solanki}, S.~K. 2020, \aap,
  639, A106

\bibitem[{{Louis} {et~al.}(2012){Louis}, {Mathew}, {Bellot Rubio}, {Ichimoto},
  {Ravindra}, \& {Raja Bayanna}}]{2012ApJ...752..109L}
{Louis}, R.~E., {Mathew}, S.~K., {Bellot Rubio}, L.~R., {et~al.} 2012, \apj,
  752, 109

\bibitem[{{Muller}(1973)}]{1973SoPh...29...55M}
{Muller}, R. 1973, \solphys, 29, 55

\bibitem[{{Ortiz} {et~al.}(2010){Ortiz}, {Bellot Rubio}, \& {Rouppe van der
  Voort}}]{2010ApJ...713.1282O}
{Ortiz}, A., {Bellot Rubio}, L.~R., \& {Rouppe van der Voort}, L. 2010, \apj,
  713, 1282

\bibitem[{{Parker}(1979)}]{1979ApJ...234..333P}
{Parker}, E.~N. 1979, \apj, 234, 333

\bibitem[{{Rempel}(2011)}]{2011ApJ...729....5R}
{Rempel}, M. 2011, \apj, 729, 5

\bibitem[{{Rempel} \& {Cheung}(2014)}]{2014ApJ...785...90R}
{Rempel}, M. \& {Cheung}, M.~C.~M. 2014, \apj, 785, 90

\bibitem[{{Rempel} {et~al.}(2009){Rempel}, {Sch{\"u}ssler}, \&
  {Kn{\"o}lker}}]{2009ApJ...691..640R}
{Rempel}, M., {Sch{\"u}ssler}, M., \& {Kn{\"o}lker}, M. 2009, \apj, 691, 640

\bibitem[{{Riethm{\"u}ller} {et~al.}(2008){Riethm{\"u}ller}, {Solanki}, \&
  {Lagg}}]{2008ApJ...678L.157R}
{Riethm{\"u}ller}, T.~L., {Solanki}, S.~K., \& {Lagg}, A. 2008, \apjl, 678,
  L157

\bibitem[{{Riethm{\"u}ller} {et~al.}(2013){Riethm{\"u}ller}, {Solanki}, {van
  Noort}, \& {Tiwari}}]{2013A&A...554A..53R}
{Riethm{\"u}ller}, T.~L., {Solanki}, S.~K., {van Noort}, M., \& {Tiwari}, S.~K.
  2013, \aap, 554, A53

\bibitem[{{Rimmele}(2008)}]{2008ApJ...672..684R}
{Rimmele}, T. 2008, \apj, 672, 684

\bibitem[{{Scharmer} \& {Henriques}(2012)}]{2012A&A...540A..19S}
{Scharmer}, G.~B. \& {Henriques}, V.~M.~J. 2012, \aap, 540, A19

\bibitem[{{Scharmer} {et~al.}(2011){Scharmer}, {Henriques}, {Kiselman}, \& {de
  la Cruz Rodr{\'\i}guez}}]{2011Sci...333..316S}
{Scharmer}, G.~B., {Henriques}, V.~M.~J., {Kiselman}, D., \& {de la Cruz
  Rodr{\'\i}guez}, J. 2011, Science, 333, 316

\bibitem[{{Scharmer} {et~al.}(2008){Scharmer}, {Nordlund}, \&
  {Heinemann}}]{2008ApJ...677L.149S}
{Scharmer}, G.~B., {Nordlund}, {\AA}., \& {Heinemann}, T. 2008, \apjl, 677,
  L149

\bibitem[{{Sch{\"u}ssler} \& {V{\"o}gler}(2006)}]{2006ApJ...641L..73S}
{Sch{\"u}ssler}, M. \& {V{\"o}gler}, A. 2006, \apjl, 641, L73

\bibitem[{{Siu-Tapia} {et~al.}(2017){Siu-Tapia}, {Lagg}, {Solanki}, {van
  Noort}, \& {Jur{\v{c}}{\'a}k}}]{2017A&A...607A..36S}
{Siu-Tapia}, A., {Lagg}, A., {Solanki}, S.~K., {van Noort}, M., \&
  {Jur{\v{c}}{\'a}k}, J. 2017, \aap, 607, A36

\bibitem[{{Sobotka} {et~al.}(1997){Sobotka}, {Brandt}, \&
  {Simon}}]{1997A&A...328..682S}
{Sobotka}, M., {Brandt}, P.~N., \& {Simon}, G.~W. 1997, \aap, 328, 682

\bibitem[{{Sobotka} \& {Jur{\v{c}}{\'a}k}(2009)}]{2009ApJ...694.1080S}
{Sobotka}, M. \& {Jur{\v{c}}{\'a}k}, J. 2009, \apj, 694, 1080

\bibitem[{{Sobotka} \& {Puschmann}(2009)}]{2009A&A...504..575S}
{Sobotka}, M. \& {Puschmann}, K.~G. 2009, \aap, 504, 575

\bibitem[{{Socas-Navarro} {et~al.}(2004){Socas-Navarro}, {Mart{\'\i}nez
  Pillet}, {Sobotka}, \& {V{\'a}zquez}}]{2004ApJ...614..448S}
{Socas-Navarro}, H., {Mart{\'\i}nez Pillet}, V., {Sobotka}, M., \&
  {V{\'a}zquez}, M. 2004, \apj, 614, 448

\bibitem[{{Solanki} \& {Montavon}(1993)}]{1993A&A...275..283S}
{Solanki}, S.~K. \& {Montavon}, C.~A.~P. 1993, \aap, 275, 283

\bibitem[{{Title} {et~al.}(1993){Title}, {Frank}, {Shine}, {Tarbell}, {Topka},
  {Scharmer}, \& {Schmidt}}]{1993ApJ...403..780T}
{Title}, A.~M., {Frank}, Z.~A., {Shine}, R.~A., {et~al.} 1993, \apj, 403, 780

\bibitem[{{Tiwari} {et~al.}(2013){Tiwari}, {van Noort}, {Lagg}, \&
  {Solanki}}]{2013A&A...557A..25T}
{Tiwari}, S.~K., {van Noort}, M., {Lagg}, A., \& {Solanki}, S.~K. 2013, \aap,
  557, A25

\bibitem[{{T\"onjes} \& {W\"ohl}(1982)}]{1982SoPh...75...63T}
{T\"onjes}, K. \& {W\"ohl}, H. 1982, \solphys, 75, 63

\bibitem[{{Tsuneta} {et~al.}(2008){Tsuneta}, {Ichimoto}, {Katsukawa}, {Nagata},
  {Otsubo}, {Shimizu}, {Suematsu}, {Nakagiri}, {Noguchi}, {Tarbell}, {Title},
  {Shine}, {Rosenberg}, {Hoffmann}, {Jurcevich}, {Kushner}, {Levay}, {Lites},
  {Elmore}, {Matsushita}, {Kawaguchi}, {Saito}, {Mikami}, {Hill}, \&
  {Owens}}]{2008SoPh..249..167T}
{Tsuneta}, S., {Ichimoto}, K., {Katsukawa}, Y., {et~al.} 2008, \solphys, 249,
  167

\bibitem[{{van Noort}(2012)}]{2012A&A...548A...5V}
{van Noort}, M. 2012, \aap, 548, A5

\bibitem[{{van Noort} {et~al.}(2013){van Noort}, {Lagg}, {Tiwari}, \&
  {Solanki}}]{2013A&A...557A..24V}
{van Noort}, M., {Lagg}, A., {Tiwari}, S.~K., \& {Solanki}, S.~K. 2013, \aap,
  557, A24

\bibitem[{{Watanabe}(2014)}]{2014PASJ...66S...1W}
{Watanabe}, H. 2014, \pasj, 66, S1

\bibitem[{{Watanabe} {et~al.}(2012){Watanabe}, {Bellot Rubio}, {de la Cruz
  Rodr{\'\i}guez}, \& {Rouppe van der Voort}}]{2012ApJ...757...49W}
{Watanabe}, H., {Bellot Rubio}, L.~R., {de la Cruz Rodr{\'\i}guez}, J., \&
  {Rouppe van der Voort}, L. 2012, \apj, 757, 49

\bibitem[{{Watanabe} {et~al.}(2009){Watanabe}, {Kitai}, \&
  {Ichimoto}}]{2009ApJ...702.1048W}
{Watanabe}, H., {Kitai}, R., \& {Ichimoto}, K. 2009, \apj, 702, 1048

\bibitem[{{Zakharov} {et~al.}(2008){Zakharov}, {Hirzberger}, {Riethm{\"u}ller},
  {Solanki}, \& {Kobel}}]{2008A&A...488L..17Z}
{Zakharov}, V., {Hirzberger}, J., {Riethm{\"u}ller}, T.~L., {Solanki}, S.~K.,
  \& {Kobel}, P. 2008, \aap, 488, L17

\end{thebibliography}

\end{document}